\documentclass[11pt]{article}
\usepackage{graphicx}
\usepackage[intlimits]{amsmath}
\usepackage{amsfonts}
\usepackage{amssymb}
\usepackage{fancyhdr}
\usepackage{epsfig}
\usepackage{rotating}
\usepackage{lscape}
\usepackage{graphicx}

\def\B      {\ensuremath{B}\hbox{ }}
\def\Bp      {\ensuremath{B^{+}}\hbox{ }}

\newcommand {\bpiz} {\ensuremath{B^{+} \rightarrow \piz \ell^+ \nu}\xspace}
\newcommand {\bpi} {\ensuremath{B^0 \rightarrow \pim \ell^+ \nu}\xspace}
\newcommand {\bpigen} {\ensuremath{B \rightarrow \pi \ell \nu}\xspace}
\newcommand {\brhogen} {\ensuremath{B \rightarrow \rho \ell \nu}\xspace}

\newcommand {\bulnu}{\ensuremath{\b \rightarrow \u \ell \nu}\hbox{ }}

\newcommand {\Bxulnu}{\ensuremath{\B \rightarrow X_u \ell \nu}\xspace}
\newcommand {\Bxlnu}{\ensuremath{\B \rightarrow X \ell \nu}\xspace}
\newcommand {\Bnxlnu}{\ensuremath{\B^{0} \rightarrow X \ell \nu}}
\newcommand {\Bpxlnu}{\ensuremath{\B^{\pm} \rightarrow X \ell \nu}}

\newcommand {\mmiss}{\ensuremath{m_{miss}^2}\xspace}

\newcommand {\breco}{\ensuremath{B_{reco}}\xspace}

\newcommand {\bsig}{\ensuremath{B_{sig}}\xspace}

\newcommand {\D}{\ensuremath{D}\xspace}
\newcommand{\beq}{\begin{equation}}
\newcommand{\beqa}{\begin{eqnarray}}
\newcommand{\beqn}{\begin{eqnarray}}
\newcommand{\eeq}{\end{equation}}
\newcommand{\eeqa}{\end{eqnarray}}
\newcommand{\eeqn}{\end{eqnarray}}
\makeatletter
\def\slash#1{{\mathpalette\c@ncel{#1}}} 
\makeatother

\def\etal               {{\it et~al.,}}

\newcommand{\BABARPubYear}    {05}

\newcommand{\BABARConfNumber} {17}
\newcommand{\SLACPubNumber} {11369}
\newcommand{\LANLNumber}   {0507085}

\input babarsym

\long\def\inst#1{\par\nobreak\kern 4pt\nobreak
    {\it #1}\par\vskip 10pt plus 3pt minus 3pt}

\begin{document}

{\pagestyle{empty}


\begin{flushright}
\babar-CONF-\BABARPubYear/\BABARConfNumber \\
SLAC-PUB-\SLACPubNumber \\
hep-ex/\LANLNumber \\
July 2005
\end{flushright}

\par\vskip 1.2cm

\begin{center}
\Large \bf Measurement of the \bpi  and \bpiz Branching Fractions
and Determination of \Vub in $\FourS\ra\BB$ Events\\
Tagged by a Fully Reconstructed \B Meson
\end{center}
\bigskip

\begin{center}
\large The \babar\ Collaboration\\
\mbox{ }\\
\today
\end{center}
\bigskip \bigskip

\begin{center}
\large \bf Abstract
\end{center}
We report preliminary measurements of the charmless exclusive semileptonic branching fractions 
of the \bpi and \bpiz decays, based on $211 \invfb$ of data collected at the \FourS resonance
by the \babar\ detector. 
In events in which the decay of one \B meson to a hadronic final state is fully reconstructed, the 
semileptonic decay of the second $B$ meson is identified by the detection of a charged lepton and a pion. 
We measure the partial branching fractions for \bpi and \bpiz in three regions of the invariant mass squared
of the lepton pair, and we obtain the total branching fractions 
$\BR(\bpi) = (1.14 \pm 0.27_{stat} \pm 0.17_{syst})\times 10^{-4}$ and $\BR(\bpiz) =
(0.86 \pm 0.22_{stat} \pm 0.11_{syst})\times 10^{-4}$. Using isospin symmetry, we measure the
combined total branching fraction  
$\BR(\bpi) = (1.28 \pm 0.23_{stat} \pm 0.16_{syst})\times 10^{-4}$.
Theoretical predictions of the form-factor are used to determine the magnitude of the 
Cabibbo-Kobayashi-Maskawa matrix element 
$\Vub = (3.7 \pm 0.3_{stat} \pm 0.2_{syst} {{}^{+0.8}_{-0.5}}_{FF})\times10^{-3}$, 
 where the last uncertainty is due to the form-factor normalization.
\vfill
\begin{center}
Submitted to the International Europhysics Conference On High-Energy Physics (HEP 2005),
7/21---7/27/2005, Lisbon, Portugal
\end{center}

\vspace{1.0cm}
\begin{center}
{\em Stanford Linear Accelerator Center, Stanford University, 
Stanford, CA 94309} \\ \vspace{0.1cm}\hrule\vspace{0.1cm}
Work supported in part by Department of Energy contract DE-AC03-76SF00515.
\end{center}

\newpage
} 

\begin{center}
\small

The \babar\ Collaboration,
\bigskip

B.~Aubert,
R.~Barate,
D.~Boutigny,
F.~Couderc,
Y.~Karyotakis,
J.~P.~Lees,
V.~Poireau,
V.~Tisserand,
A.~Zghiche
\inst{Laboratoire de Physique des Particules, F-74941 Annecy-le-Vieux, France }
E.~Grauges
\inst{IFAE, Universitat Autonoma de Barcelona, E-08193 Bellaterra, Barcelona, Spain }
% ...

\end{center}\newpage

\newpage
\clearpage

\section{Introduction}
Precise measurements of the Cabibbo-Kobayashi-Maskawa matrix element $V_{ub}$
can be employed to test the consistency of the Standard Model description of $CP$ violation.
\Vub can be extracted from charmless semileptonic \B decays, such as \bpigen 
\footnote{Charge-conjugate modes are implied throughout this paper, unless explicitly stated otherwise.} 
and $B\ra \rho\ell\nu$,
allowing for more stringent kinematic constraints and better background suppression 
than possible with inclusive measurements. 

However the determination of \Vub from exclusive decays is complicated by the presence 
of the strong interaction between the quarks in the initial and 
the final states. In the case of \bpigen decays 
the dynamics are described by a single 
form-factor $f(q^2)$ that depends on the square of the $B\to\pi$ momentum transfer. 
The shape of the form-factors can in principle be measured, while for 
their normalization we have to rely on theoretical predictions \cite{Ball05,
HPQCD04,FNAL04}.
The measurements presented here rely on these theoretical predictions for both the shape and 
the normalization.

We present a measurement of the branching fractions of the exclusive charmless semileptonic 
decays \bpi and \bpiz.  The analysis is based on a sample of \BB events produced at the \FourS
resonance that are tagged by a fully reconstructed hadronic decay.
A semileptonic decay of the recoiling \B meson is identified by the
presence of a charged lepton. 
The charmless mesons in the semileptonic decay are reconstructed 
and the missing mass is calculated assuming that the pion and the charged lepton are 
the only particles present in the recoil except for the undetected neutrino.
Since the momentum of the tagging \B meson is measured, a transformation to the rest frame 
of the recoiling $B$ meson can be applied. The full reconstruction approach provides also a very clean sample of \BB  
events, determines the flavor of the reconstructed \B meson and  
separates \Bz and \Bp decays. 

Exclusive charmless semileptonic \B decays have been previously measured by
the CLEO \cite{CLEOpilnu}, Belle \cite{Bellepilnu} and BaBar \cite{bad1158,piplnu,pizlnu} 
collaborations. The approach presented in this paper results in small backgrounds in 
the whole $q^2$ spectrum with very loose kinematic selection criteria. 
We present here the partial branching fractions in three $q^2$ regions, $q^2 < 8 \gev^2/c^4$, $8 < q^2 < 16
\gev^2/c^4$, $q^2 > 16 \gev^2/c^4$, 
 the measurement of the total branching fraction, and the measurement of \Vub.

\section{Data Sample}

The preliminary results are based on a data sample of about 233 million \BB pairs, corresponding to 211 \invfb, 
collected with the \babar\ detector \cite{babar} at
 the SLAC \pep2\ asymmetric-energy $e^+e^-$ collider \cite{pep2}, operating at the \FourS resonance.
A Monte Carlo (MC) simulation of the \babar\ detector based on
Geant4 \cite{geant} has been used to optimize the selection criteria and to
determine the signal efficiencies and background distributions. 

\section{Overview of Analysis Method}
\label{sec:strategy}

The analysis strategy follows closely that of Ref. \cite{ichep04}. 
The starting point is the selection of a sample of events
in which the hadronic decay of one of the two \B mesons (\breco) is fully 
reconstructed.
About 1000 different $B \rightarrow D Y$ decay chains are selected, 
where $D$ refers to a 
charm meson and $Y$ represents a collection of hadrons with total charge 
$\pm 1$ and composed of $n_1 \pi^{\pm} + n_2 K^{\pm}  + n_3 \KS + n_4\piz$,
with $n_1 + n_2 < 6$,  $n_3 < 3$ and $n_4 <  3$. 

We reconstruct $D^-\ra K^+\pi^-\pi^-$, $K^+\pi^-\pi^-\piz$, $\KS\pi^-$,
$\KS\pi^-\piz$, $\KS\pi^-\pi^-\pi^+$; 
$D^{*-}\ra\Dzb\pi^-$; 
$\Dzb\ra K^+\pi^-$,
$K^+\pi^-\piz$, $K^+\pi^-\pi^-\pi^+$,  $\KS\pi^+\pi^-$;
and $\Dstarzb \ra\Dzb\piz, \Dzb\gamma$.
Then we use $D^-$  and $D^{*-}$ ($\Dzb$ and  $\Dstarzb$) decays as a ``seed'' to
reconstruct $B^0$ ($B^+$) decays.
Overall, we correctly reconstruct one $B$ candidate in 
0.3\%  (0.5\%) of the \BzBzb\ (\BpBm) events.
The kinematic consistency of a \breco candidate with a \B meson decay is 
checked using two variables: the beam-energy substituted mass
$\mes = \sqrt{s/4 - \vec{p}^{\,2}_B}$
and the energy difference $\Delta E  = E_B  - \sqrt{s}/2$.
Here  $\sqrt{s}$ refers to  the total
energy in the \FourS center of  mass frame, and $\vec{p}_{B}$ and 
$E_B$ denote
the momentum and energy of the \breco\ candidate in the same frame. 
For signal events the \mes\ and $\Delta E $ distributions peak at the \B meson mass
and at zero, respectively.

\begin{figure}[!b]
 \begin{centering}
 \epsfig{file=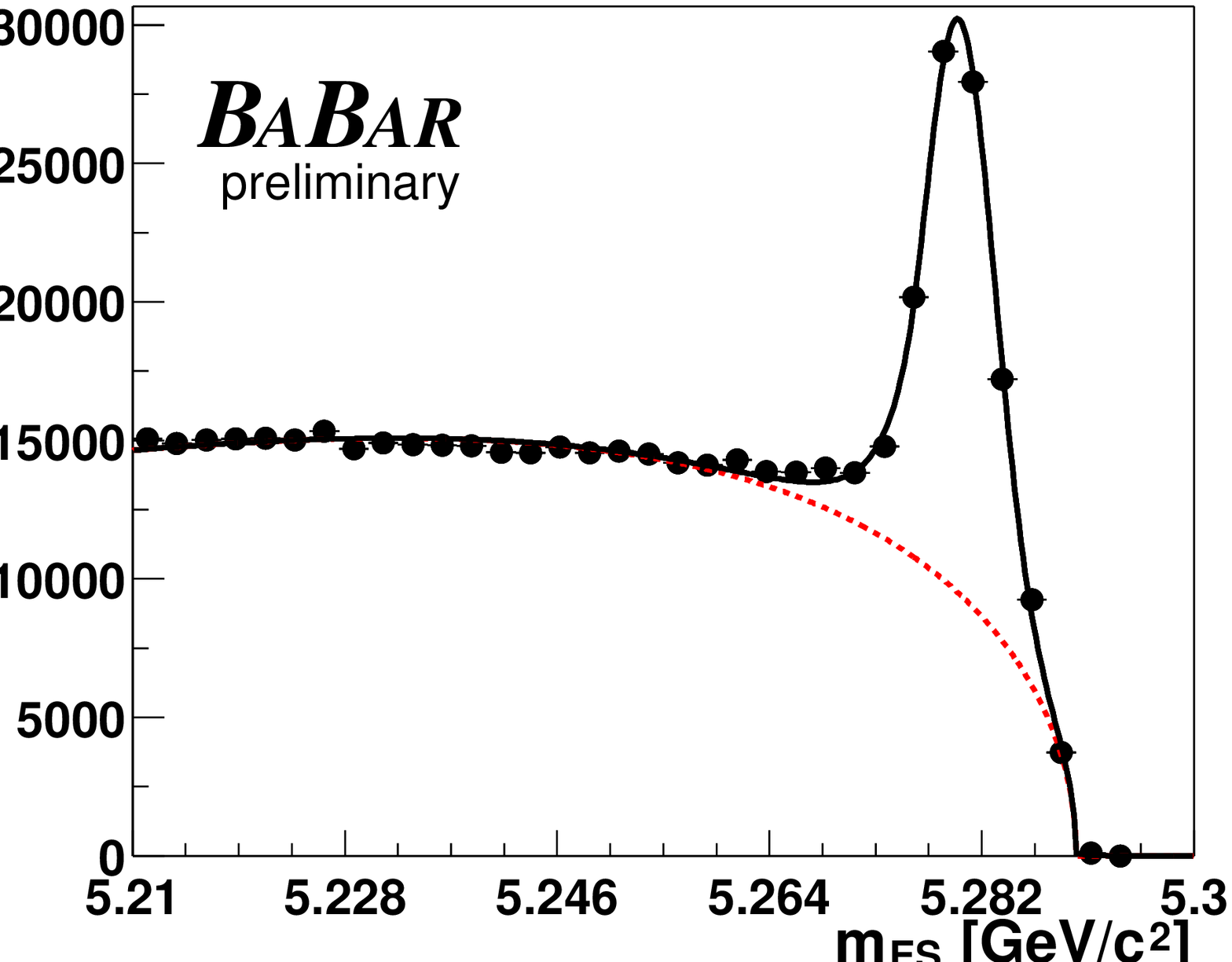,width=6.7cm}
 \epsfig{file=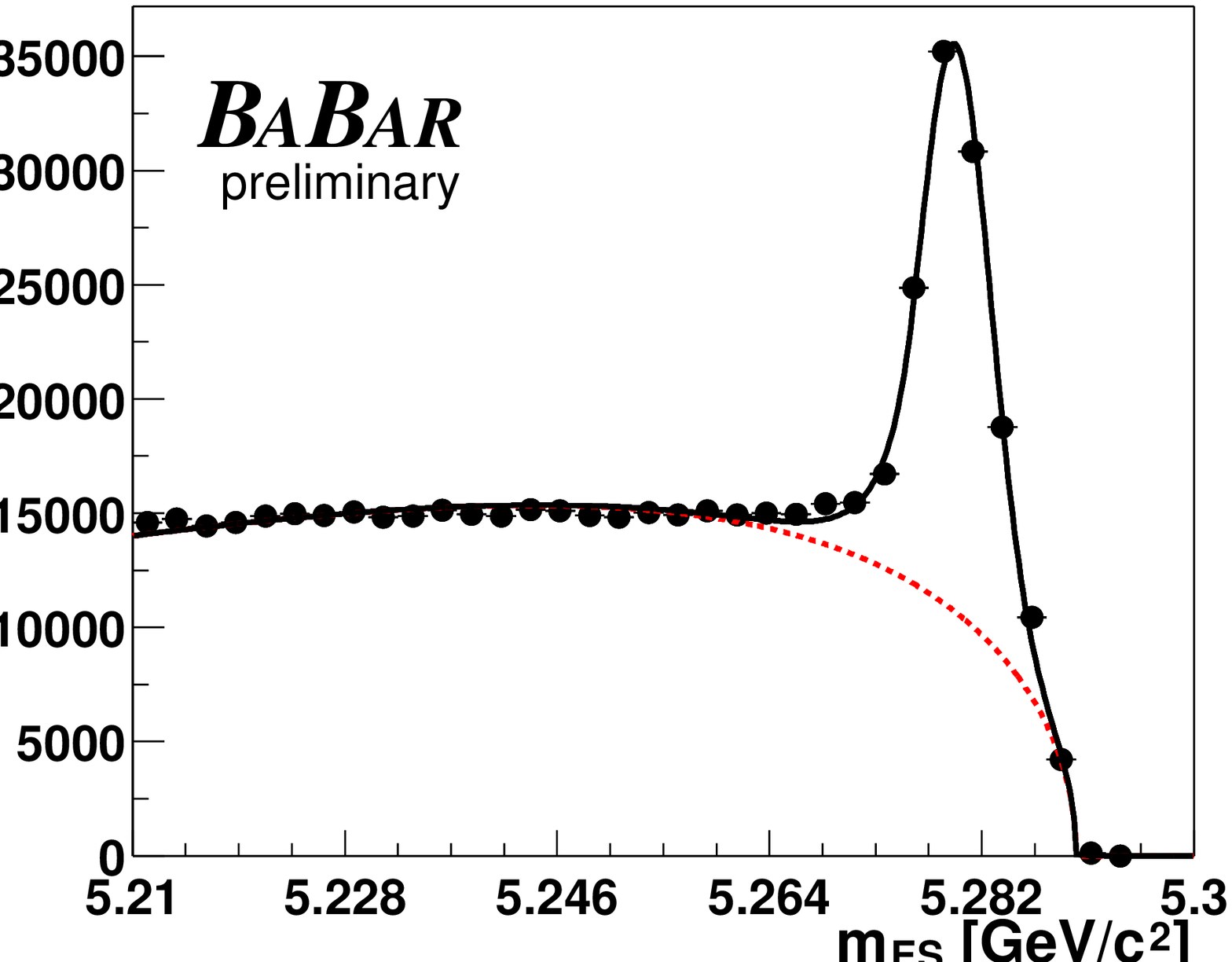,width=6.7cm}
 \epsfig{file=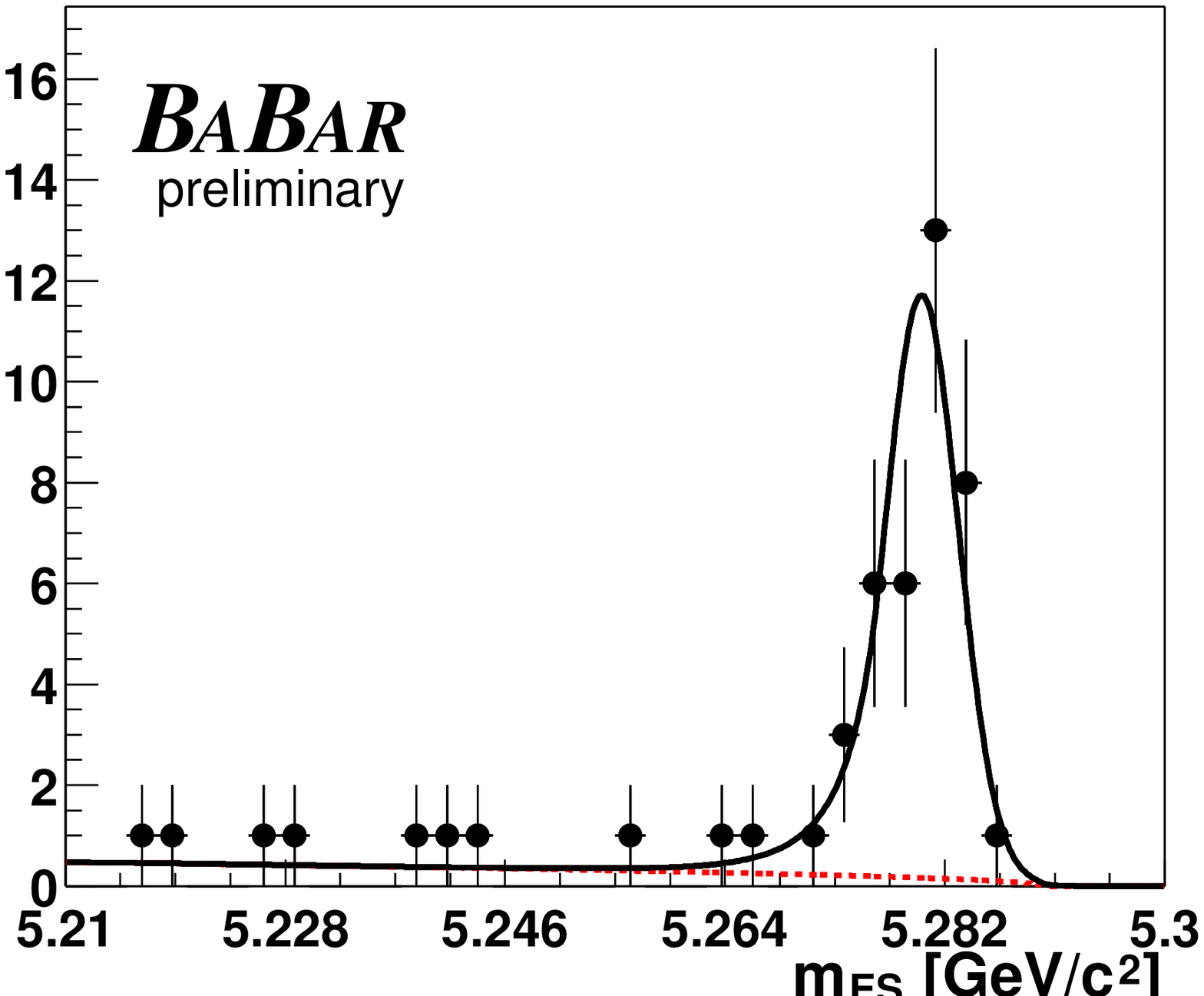,width=6.7cm}
 \epsfig{file=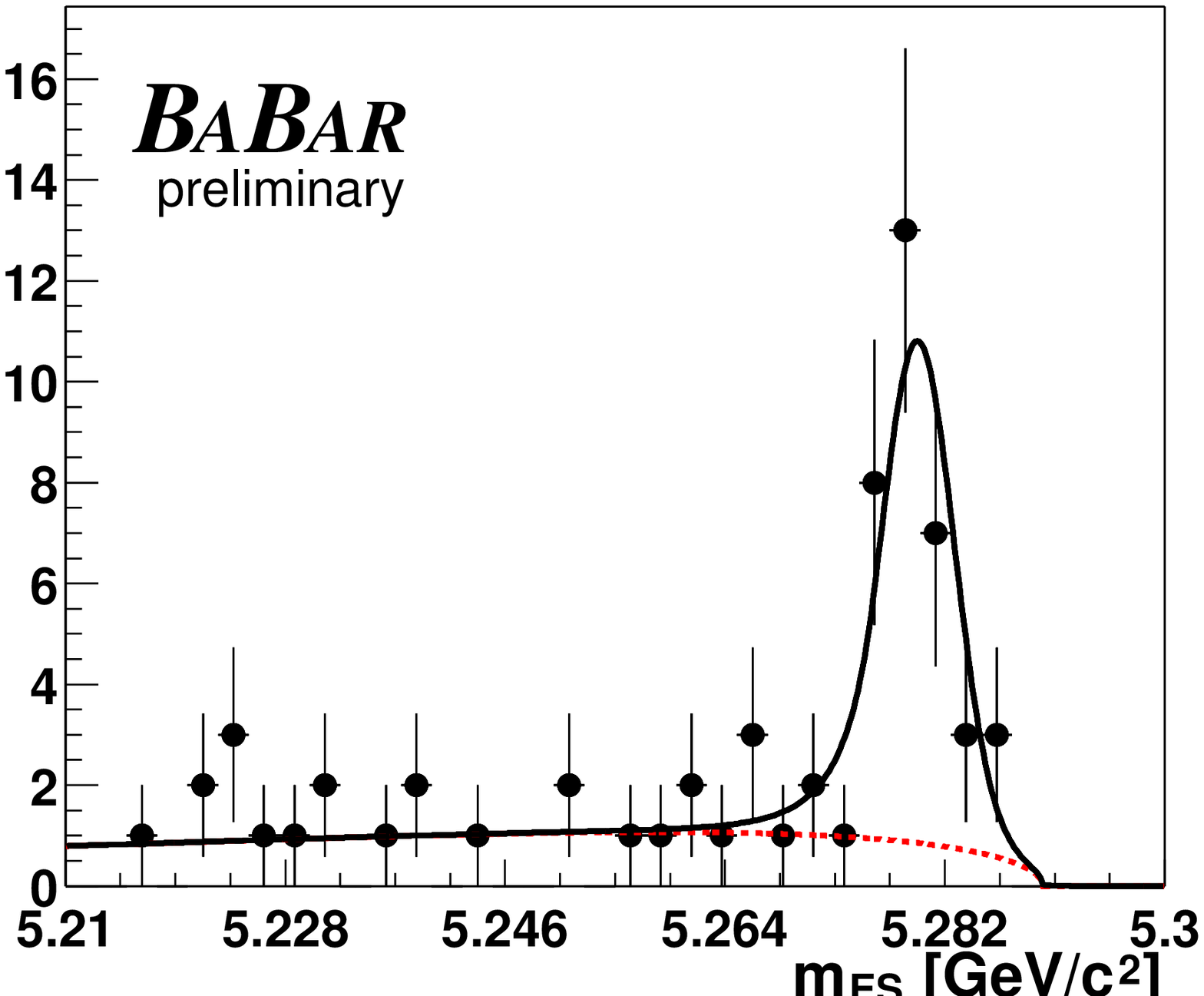,width=6.7cm}
 \caption{ Fit to the \mes distribution for events with a fully reconstructed \Bz (left) or \Bp (right) decay, after the request of a prompt energetic lepton (top) and after all selection criteria for \bpigen decays (bottom). The fitted curve (black line) to the data points (black dots) is the sum of a radiation loss corrected gaussian and a threshold function described by Eq.~\ref{eq:argus} (dashed line).
\label{fig:mesfit}}
 \end{centering}
\end{figure}

The combinatorial background from \BB\ events and  
$e^+e^-\to q\bar{q}$ ($q=u,d,s,c$) production, in the \breco\ sample, is subtracted by performing an unbinned likelihood fit
to the \mes\ distribution, using the following threshold function \cite{argus}

\begin{equation}
\label{eq:argus}
\frac{dN}{d\mes} = N\cdot \mes \cdot \sqrt{1-x^2} \cdot \exp \left(- \xi \cdot (1 - x^2)\right)\,
\end{equation}

\noindent for the background (where $x = \mes/m_{\rm max}$ and $m_{\rm max}$ is the endpoint of the curve)
and a Gaussian function corrected for radiation losses
\cite{crystallball} peaked at the \B meson mass for the signal (Fig. \ref{fig:mesfit}).

Once a \breco meson has been reconstructed in the event, the selection of
the other \B meson (\bsig) decay follows in two steps. First, a ``semileptonic selection'' is applied to obtain
a sample of semileptonic decays. Then, a refined selection is applied  to
build a sample of $\bpigen$ decays, as described in Sec.~\ref{sec:selection}. In order to minimize the systematic uncertainties, the exclusive branching fractions are measured relative to the inclusive semileptonic one.

The semileptonic selection is based essentially on the presence of a charged (electron or muon) lepton 
with its momentum $p^{*}_{lep}$ in the \bsig rest frame
greater than a given $p^{*}_{cut}$. The number of \Bxlnu\ events, $N_{sl}^{meas}$, and the number of 
remaining background events, $BG_{sl}$, peaking at the \B mass in the \mes distribution and estimated with the Monte Carlo
simulation, are related to the true number of semileptonic
decays $N_{sl}^{true}$ as
\begin{equation}
N_{sl}^{meas} - BG_{sl} = \epsilon_l^{sl} \epsilon_t^{sl}N_{sl}^{true}.
\end{equation}
Here $\epsilon_l^{sl}$ refers to the efficiency for selecting a lepton
from a semileptonic \B decay with a momentum above $p^{*}_{cut}$ in an
event with a hadronic \B decay, reconstructed with tag efficiency $\epsilon_t^{sl}$. 
The background normalization is 
taken from the Monte Carlo scaled to the luminosity of the data sample.

The full reconstruction of the \breco meson allows for a precise determination 
of the neutrino four-momentum $p_{\nu}$,
estimated from the missing four-momentum in the event: 

\begin{equation}
\label{eq:pmiss}
p_\nu = p_{miss} = p_{\Upsilon(4S)} - p_{\breco} - p_{\pi} - p_\ell, 
\end{equation}

\noindent where all momenta are measured in the laboratory frame: $p_{\Upsilon(4S)}$ is the sum of 
the four-momenta of the colliding beams, $p_{\breco}$ is the measured four-momentum of the \breco, 
$p_{\pi}$ is the measured four-momentum of the $\pi^-$ or $\piz$ and $p_{\ell}$ is the
measured four-momentum of the lepton. For signal events the only missing particle should 
be a single undetected neutrino, while for background events the missing momentum and energy in the event
are due to other undetected or poorly measured particles.
Then, in signal events the resulting missing mass, defined as $\mmiss$ = $p_{miss}^2$, peaks at zero 
and for background events 
it tends to have larger values, and allows for a powerful discrimination of
signal and background.

The number of $\bpigen$ events after the combinatoric subtraction
in a given $q^2$ range, $N_{excl}^{meas}$, and the number of peaking background events, $BG_{excl}$, 
are related to the true number of 
$\bpigen$ events $N_{excl}^{true}$ as
\begin{equation} 
N_{excl}^{meas} - BG_{excl} = \epsilon_{sel}^{excl}  \epsilon_l^{excl}
 \epsilon_t^{excl} N_{excl}^{true} ,
\label{eq:NBGexcl}
\end{equation}
\noindent
where the signal efficiency $\epsilon_{sel}^{excl}$ accounts for all selection
criteria applied to the sample after the requirement of 
a charged lepton with momentum $p^*_{lep} > p^*_{cut}$.  

The ratio between the partial branching fractions of the signal in a particular $q^2$ region 
and the branching fraction of $\Bxlnu$ decays is 
 \begin{equation}
R_{excl/sl}=
\frac{\Delta\BR(\bpigen)}{\BR(\Bxlnu)}=
 \frac{N_{excl}^{true}}{N_{sl}^{true}} = 
 \frac{(N_{excl}^{meas}- BG_{excl})/(\epsilon_{sel}^{excl})}
{(N_{sl}^{meas}-BG_{sl})} 
 \times \frac{\epsilon_l^{sl} \epsilon_t^{sl} } 
{\epsilon_l^{excl} \epsilon_t^{excl} }.
 \label{eq:ratioBR}
 \end{equation}

\noindent The ratio of efficiencies for $\Bxlnu$ and signal 
events in Eq. \ref{eq:ratioBR} is expected to be close to, but not equal to
unity.  Due to the difference in multiplicity and the different
lepton momentum spectra, we expect the tag efficiencies $\epsilon_t^{sl,excl}$
and the charged lepton efficiencies $\epsilon_l^{sl,excl}$ to be slightly different for the
two classes of events.
Finally using the semileptonic branching ratio $\BR(\Bxlnu) = (10.73 \pm 0.28)\%$ \cite{PDG2004}
and the ratio of the \Bz and \Bp lifetimes $\tau_{B^+}/\tau_{B^0}=1.086 \pm 0.017$ 
\cite{PDG2004}, the partial branching ratios $\Delta\BR(\bpi)$ and $\Delta\BR(\bpiz)$ can be derived.\\

\section{ Event Reconstruction and Selection}
\label{sec:selection}
On the recoil of a fully reconstructed \B, the \bpigen decay of \bsig\ 
is constructed by combining a pion with a charged lepton from
 only the tracks and neutral clusters which do not contribute to the \breco. 
Electron candidates are identified by a likelihood-based algorithm,
while the muon identification is based on a tight selection.
Remaining tracks are assumed to be pions if they are not identified as either a 
muon or an electron.

Neutral pions are reconstructed by using pairs of photons with an energy greater than 80 \mev 
in the laboratory. Moreover the energy of the most energetic photon used to reconstruct the \piz 
is required to be greater than 300 \mev in the \bsig\ rest frame, to reject 
combinatorial background.

In the semileptonic selection, we require:
\begin{itemize}

\item the presence of a fully reconstructed $\B_{reco}$, neutral for \bpi and 
charged for \bpiz.

\item one lepton, either an electron with $p^*_{el} > 0.5 \gevc$ or a muon with $p^*_{\mu} > 0.8 \gevc$,
originating from the \bsig.

\item correlation between the lepton charge and $\B_{reco}$ flavor for \bpiz. 
In unmixed signal events $Q_{b(reco)} Q_{\ell} < 0$, where $Q_{b(reco)}$ is the 
charge of the $b$-quark inside the \breco and  $Q_{\ell}$ is the charge of the 
 signal lepton. In mixed \Bz\ decays or in some background events $Q_{b(reco)} Q_{\ell} > 0$.
For \bpiz decays, $Q_{b(reco)} Q_{\ell} < 0$ is therefore required.
If the reconstructed \B is neutral, both events with $Q_{b(reco)} Q_{\ell} < 0$ (right sign, ``rs'') and $Q_{b(reco)} Q_{\ell} > 0$ (wrong sign, ``ws'') are accepted and the sample is subsequently corrected for the effects of \BB mixing:

\begin{equation}
\label{eq:mixing}
 N_B =  \frac{1-\chi_d}{1-2\chi_d} N_{rs}  - \frac{\chi_d}{1-2\chi_d} N_{ws} 
\end{equation}

where $\chi_d = 0.186 \pm 0.004$ \cite{PDG2004} is the mixing parameter.

\end{itemize}

To select the decay modes of interest, the following additional selection criteria are applied:

\begin{itemize}
\item event charge conservation: $Q_{tot} = Q_{\breco} + Q_{\bsig} = 0$. This condition 
rejects preferentially $b \ra c \ell \nu$ events, since their 
higher charge multiplicity leads to higher loss of charged tracks.

\item a decay-mode dependent cut on \mmiss.

\item the only tracks allowed to be present in the recoil are 
the charged lepton and the charged pion in the case of \bpi decay.

\item a $J/\psi$ veto for the \bpi mode. $J/\psi\rightarrow \ell^+\ell^-$ decays 
introduce a background due to the mis-identification of a lepton as a
pion.
To remove this background, the lepton mass hypothesis is applied to the charged pion and the invariant mass $m_{\pi\ell}$ of
the lepton-$\pi$ pair is requested to be outside the range $3.08 < m_{\pi\ell} < 3.12 \gevcc$.  
  
\item  in the \bpi channel  the residual photon energy $E_{neutral}$ is required to be smaller than 0.45 \gev to reject $B^0\rightarrow\rho^- \ell^+ \nu$ events which constitute the main \bulnu background.

\item the \piz\ reconstructed mass is required to satisfy $110<m_{\gamma\gamma}<160 \mevcc$.

\end{itemize}

The selection criteria described above have been optimized by minimizing the statistical
error on $R_{excl/sl}$ and are summarized in Tab.~\ref{tab:cuts}. 
When more than one  \bpigen  candidate is reconstructed in the same event, the
one with \mmiss closest to zero is chosen.
The selection efficiencies $\epsilon_{sel}^{excl}$ as estimated from the Monte Carlo simulation are reported in Tab.~\ref{tab:fitresq2pi} and~\ref{tab:fitresq2pi0}.
The number of events after the semileptonic selection and after all analysis cuts are 
obtained with the \mes fit described in Section \ref{sec:strategy}. The fit results on data are 
shown in Fig. \ref{fig:mesfit}.

\begin{table}[!hp]
\begin{center}
\caption{Summary of event selection for \bpi and \bpiz. 
}
\vspace{0.1in}
\begin{tabular}{|l|l|l|} 
\hline
Selection             &  \bpi & \bpiz \\
\hline\hline
Lepton momentum       & \multicolumn{2}{c|}{$p_{el}^* > 0.5 \gevc$, $p_{\mu}^* > 0.8 \gevc$,}\\
Number of leptons     & \multicolumn{2}{c|}{$N_{lepton} = 1$,}\\
Charge conservation   & \multicolumn{2}{c|}{$ Q_{tot}= 0$,} \\
Number of tracks      & \multicolumn{2}{c|}{no additional charged tracks} \\\cline{1-3}
\hline
Charge correlation    &  $Q_{b(reco)} Q_{\ell} <0$ & mixing correction \\ 
Missing mass squared  &  $|\mmiss|<0.3 \gev^2/c^4$ & $-0.5<\mmiss<0.7 \gev^2/c^4$ \\ 
\piz mass             &  - & $110<m_{\gamma\gamma}<160 \mevcc$ \\
Neutral energy        & $ E_{neutral}<0.45 \gev$ & - \\ 
Lepton-$\pi$ mass     & $|m_{\pi\ell}-m_{J/\psi}| > 0.02 \gevcc$ & - \\ \hline
\end{tabular}
\label{tab:cuts}
\end{center}
\end{table}

\section{Measurement of Branching Fractions}
\label{sec:BRextraction}

In order to extract \Vub\ from the $B\ra\pi\ell\nu$ decay rate, 
a measurement of partial branching fractions for \bpi and \bpiz is performed in bins of the invariant-mass 
squared of the lepton-neutrino system, $q^2 = 
(p_{\Upsilon(4S)} - p_{\breco} - p_{\pi})^2$.
We consider three $q^2$ bins: $q^2 < 8 $, $8 < q^2 < 16 $ and $q^2 > 16 \gev^2/c^4$.
Since the observed resolution on $q^2$ is very good (about $0.25 \gev^2/c^4$ for \bpi and $0.50 \gev^2/c^4$ 
for \bpiz) compared to the width of the $q^2$ bins, the cross-feed among the different $q^2$ bins is small
and it is considered as background. We neglect the 
correlation among the different $q^2$ bins introduced by this cross-feed. 
The peaking background surviving the selection is estimated with a Monte Carlo 
simulation and scaled to the data yields after the requirements on the charged lepton.

\begin{figure}[top]
 \begin{centering}
 \epsfig{file=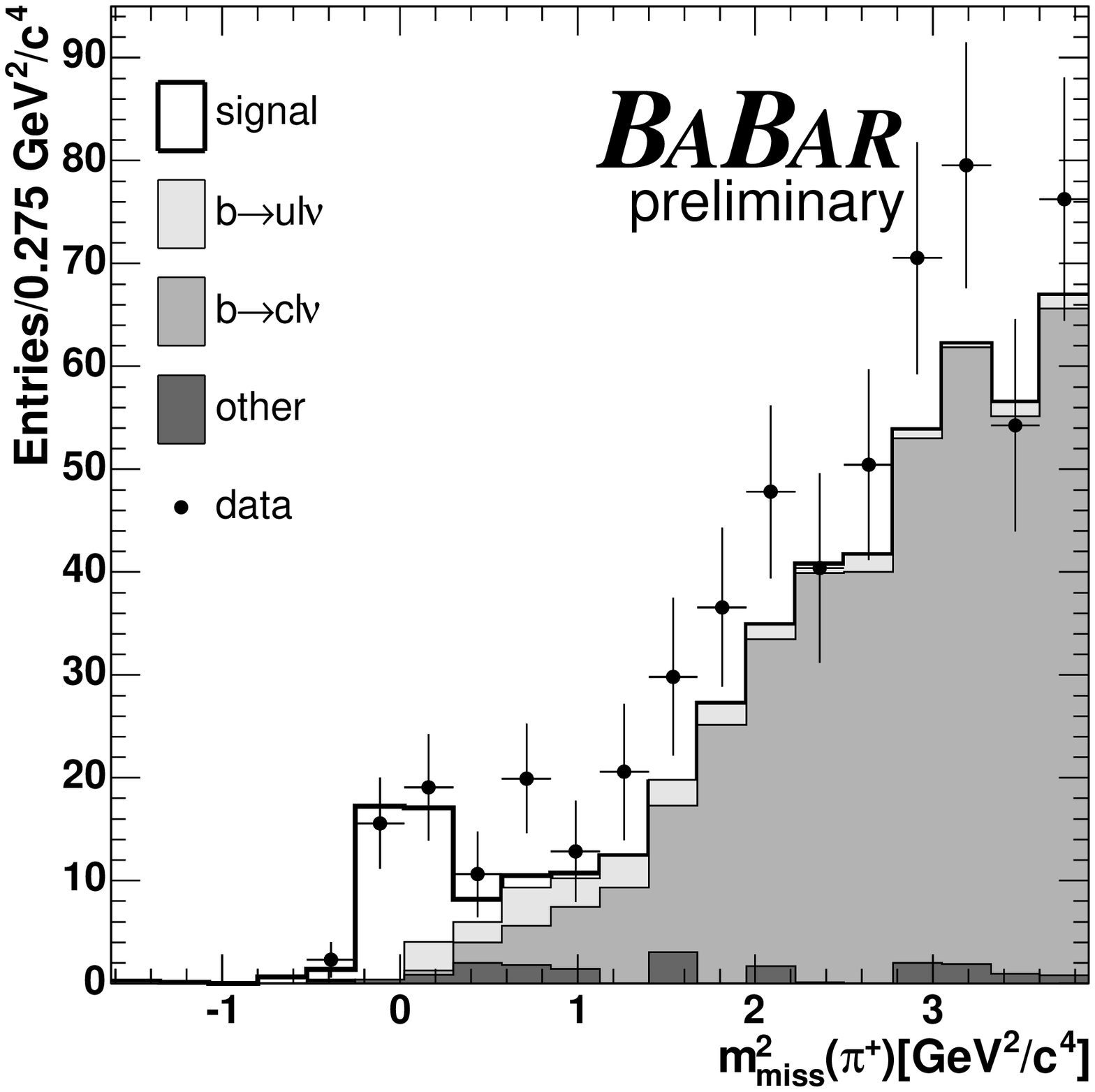,width=8.cm} 
 \caption{\mmiss distribution from data (dots) and signal and background contributions from Monte Carlo (histograms) for \bpi. The background components are scaled to the data yields after the requirements on the charged lepton.
\label{fig:datapifitnew}}
 \end{centering}
\end{figure}

\begin{figure}
 \begin{centering}
 \epsfig{file=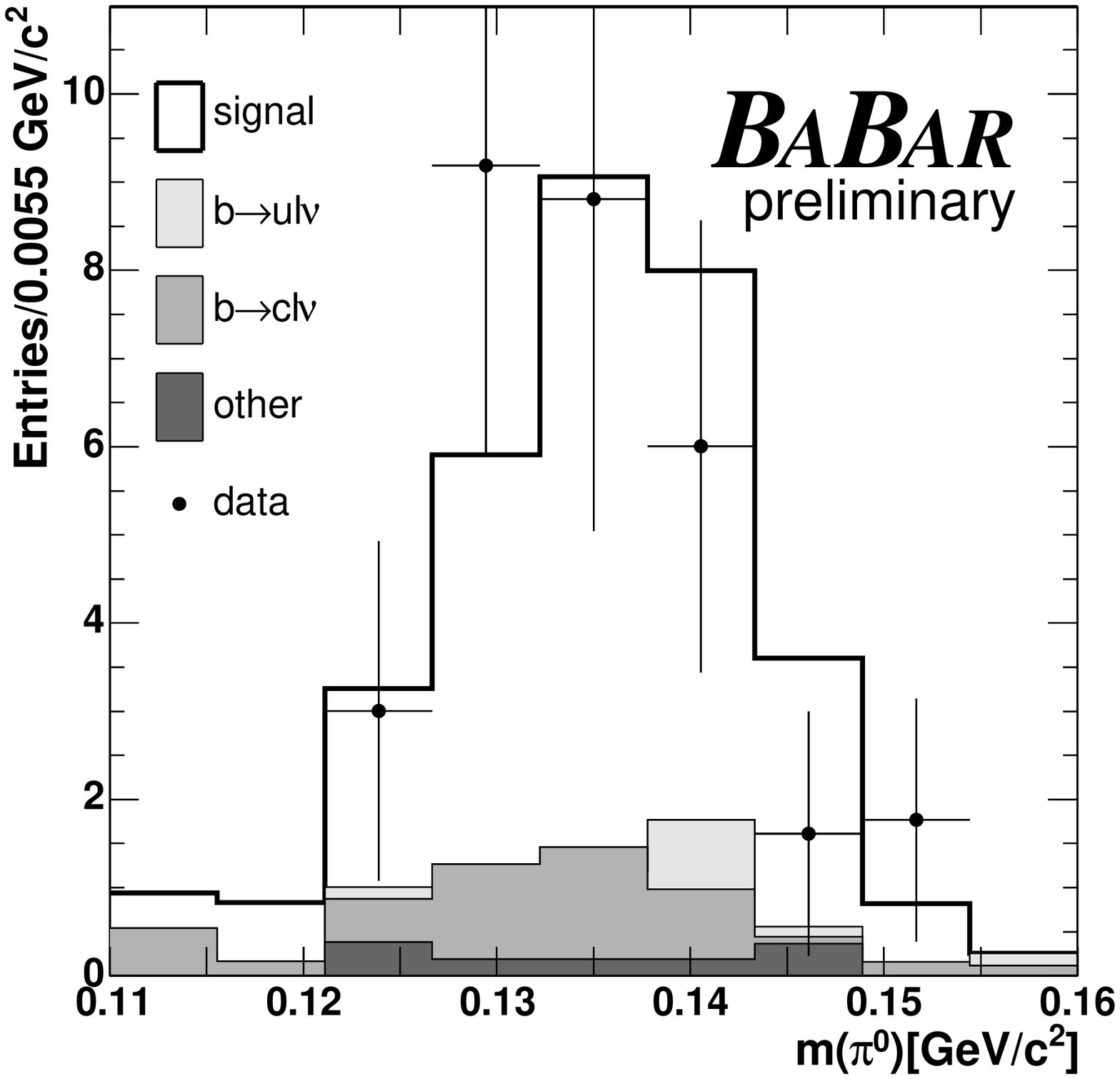,width=8.cm} 	
 \epsfig{file=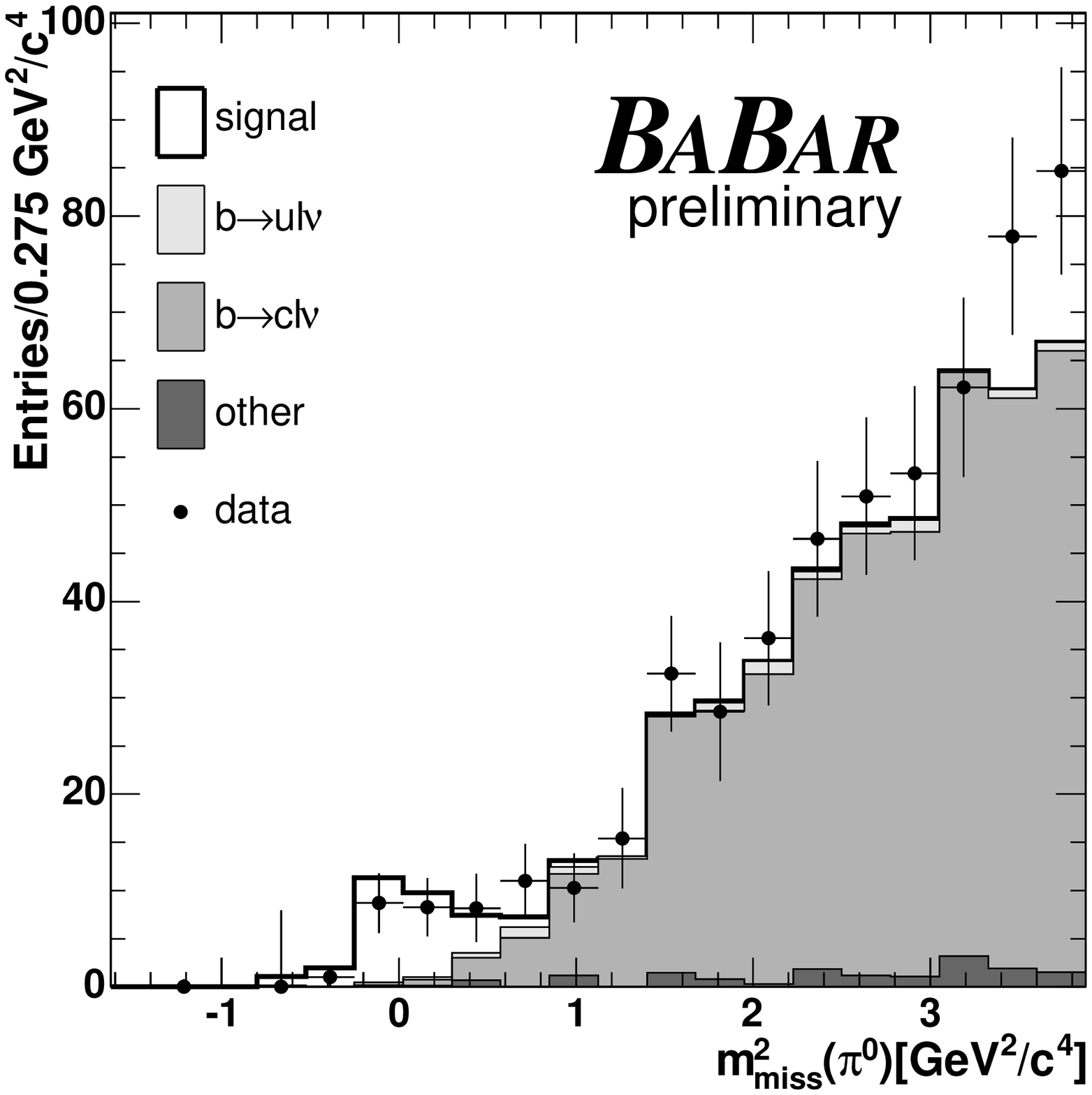,width=8.cm} 	
 \caption{$m_{\gamma\gamma}$ (left) and \mmiss (right) distributions from data (dots) and signal and background contributions from Monte Carlo (histograms)
 for \bpiz. The background components are scaled to the data yields after the requirements on the charged lepton.
\label{fig:datapi0fitnew}}
 \end{centering}
\end{figure}

The resulting partial branching fractions and all the corresponding
input measurements  are shown in Tab.~\ref{tab:fitresq2pi} and 
Tab.~\ref{tab:fitresq2pi0}, for \bpi and \bpiz respectively.
Fig. \ref{fig:datapifitnew} and \ref{fig:datapi0fitnew} show the data \mmiss and \piz mass distributions,
respectively, for events in the full $q^2$ range. Signal and background components from the Monte Carlo, 
scaled to the number of events passing the semileptonic selection, are also overlayed.

\begin{table}[h!]
\begin{center}
\caption{Measurement of  $R_{excl/sl}$ for \bpi in $q^2$ bins and corresponding inputs. The reported errors are statistical only.}
\vspace{0.1in}
\small
\begin{tabular}{|l|c|c|c|c|c|c|}
\hline
$q^2$ bin$[\gev^2/c^4]$      & $N_{excl}^{meas}$  & $BG_{excl}$  & $\epsilon_{sel}^{excl}$ & $N_{sl}^{meas}-BG_{sl}$ &$\frac{\epsilon_l^{sl} \epsilon_t^{sl} } {\epsilon_l^{excl} \epsilon_t^{excl} }$&$\frac{\Delta \BR(\bpi)}{\BR(\Bnxlnu)}[\times10^{-3}]$  \\
\hline\hline
$q^2 < 8 $            & $6.9 \pm 3.1$  & $2.7 \pm 1.6$ & $0.70 \pm 0.03$ & $43500 \pm 300$ &$ 0.97\pm 0.08 $& $0.14 \pm 0.10$ \\
$8 < q^2 < 16 $       & $10.5 \pm 3.9$ & $1.1 \pm 1.0$ & $0.54 \pm 0.03$ & $43500 \pm 300$ &$  0.86\pm 0.09   $  & $0.34 \pm 0.14$ \\
$q^2 > 16 $           & $18.7 \pm 5.0$ & $2.7 \pm 1.6$ & $0.57 \pm 0.05$ & $43500 \pm 300$ &$ 0.98\pm 0.15   $& $0.63 \pm 0.20$ \\
\hline
Total                 & $36.1 \pm 7.1$ & $6.5 \pm 2.3$ & -      & $43500 \pm 300$ & -      & $1.11 \pm 0.25$ \\
\hline
\end{tabular}
\normalsize
\label{tab:fitresq2pi}
\end{center}
\end{table}

\begin{table}[h!]
\begin{center}
\caption{Measurement of  $R_{excl/sl}$ for \bpiz in $q^2$ bins and corresponding inputs. The reported errors are statistical only.}
\vspace{0.1in}
\small
\begin{tabular}{|l|c|c|c|c|c|c|}
\hline
$q^2$ bin$[\gev^2/c^4]$    & $N_{excl}^{meas}$  & $BG_{excl}$  & $\epsilon_{sel}^{excl}$ & $N_{sl}^{meas}-BG_{sl}$ &$\frac{\epsilon_l^{sl} \epsilon_t^{sl} } {\epsilon_l^{excl} \epsilon_t^{excl} }$ &$\frac{\Delta \BR(\bpiz)}{\BR(\Bpxlnu)}[\times10^{-3}]$  \\
\hline\hline
$q^2 < 8 $            & $7.7  \pm 2.9$ & $2.6 \pm 1.6$ & $0.44 \pm 0.04$ & $69600 \pm 400$ & $0.97\pm 0.12$ &$0.16 \pm 0.09$ \\
$8 < q^2 < 16 $       & $13.5 \pm 4.0$ & $2.9 \pm 1.7$ & $0.42 \pm 0.04$ & $69600 \pm 400$ & $1.01\pm 0.09$&$0.37 \pm 0.14$ \\
$q^2 > 16 $           & $12.9 \pm 3.8$ & $4.1 \pm 2.0$ & $0.37 \pm 0.06$ & $69600 \pm 400$ & $0.72\pm 0.13 $&$0.24 \pm 0.10$ \\
\hline
Total                 & $34.1 \pm 6.2$ & $9.6 \pm 3.1$ & -    & $69600 \pm 400$& -     & $0.77 \pm 0.19$ \\
\hline
\end{tabular}
\normalsize
\label{tab:fitresq2pi0}
\end{center}
\end{table}


\section{Systematic Uncertainties}
\label{sec:syst}

Since the estimation of the systematic uncertainties is affected by low statistics in the Monte Carlo,
they have been calculated in the whole $q^2$ range and assumed to be the same in each of 
the three $q^2$ bins except for the systematic uncertainties due to form-factors and to Monte Carlo statistics.
These uncertainties have been assessed separately for each $q^2$ bin.

Uncertainties related to the reconstruction of charged tracks
are  determined  by  removing   randomly  a  fraction  of  tracks
corresponding  to  the uncertainty  in  the  track finding  efficiency.   
The  systematic  error  due  to  the
reconstruction of neutral  particles in the EMC is studied by varying the 
resolution and efficiency to match
those found in control samples in data.
We estimated the systematics due to particle  identification by
varying the  electron and muon identification efficiency  by $\pm 2\%$ and $\pm3\%$, 
respectively and the relative
mis-identification probabilities by $15\%$. 

The uncertainty of the $B_{reco}$ background subtraction
is estimated by changing the signal shape function to a Gaussian function.
We evaluated the effect of cross-feed between \Bz\ and \Bp\ decays by repeating 
the analysis with only the $B^0\bar{B}^0$ or $B^+B^-$ Monte Carlo sample for \bpi or \bpiz respectively. 
The uncertainty on the ratio of efficiencies for $\Bxlnu$ and signal events in Eq. \ref{eq:ratioBR}, 
due to limited Monte Carlo statistics, has been taken as a systematic uncertainty. 

The impact of the charm semileptonic branching fractions has been estimated
by varying each of the exclusive  branching fractions within one standard deviation  of the
current world average~\cite{PDG2004}.
The effects due to exclusive \Bxulnu\ decays have been evaluated
by varying their branching fractions by $30\%$ for $\B \ra \rho \ell \nu$,
by $40\%$ for $\B \ra \omega \ell \nu$ and by
$100\%$ for the remaining decay modes.

The use of different theoretical models changes the lepton spectrum shape for the signal and, as a 
consequence, affects the efficiencies $\epsilon_l^{excl}$, $\epsilon_l^{sl}$ and $\epsilon_{sel}^{excl}$. 
The Monte Carlo samples used in this analysis were generated using the ISGW2 model \cite{ref:isgwtwo}. 
We reweighted the event distributions according to the recent calculations 
by Ball and Zwicky \cite{Ball05} based on 
light-cone sum rules (LCSR) since, among the calculations currently available, 
these calculations result in distributions that differ most from those predicted by ISGW2. 
We assign the variations with respect to the ISGW2 as 
 systematic uncertainties.  This contribution is small because the signal 
efficiencies for \bpi and \bpiz are largely independent of $q^2$.
\newpage
\noindent A summary of the systematic uncertainties discussed above is shown in Tab. \ref{tab:systematics2}. 

\begin{table}[htbp]
\caption{Systematic uncertainties in the measurement of
$R_{excl/sl}$.}
\begin{center}
\begin{tabular}{|l|c|c|}
\hline
&\multicolumn{2}{c|}{Relative Uncertainty on $R_{excl/sl}$($\%$)} \\\cline{2-3}
                                   & \bpi & \bpiz \\
\hline
\hline
Track reconstruction efficiency                & 1.1 & 1.4 \\
Photon resolution, \piz reconstruction        & 1.2 & 3.7 \\
Electron identification                        & 1.1 & 1.1 \\
Muon identification                            & 2.3 & 2.3 \\
\mes fit                           & 9.4 & 5.0 \\
Cross-feed $\Bz \leftrightarrow \Bp$& 0.7 & 1.4 \\
\hline
$B\to Dl\nu X$ and $D$ branching fractions        & 0.2 & 2.6 \\
\Bxulnu branching fractions                    & 4.2 & 1.7 \\
\hline
Form-factor~model dependence ($q^2 < 8$)       & 0.3  & 0.3 \\
Form-factor~model dependence ($8 < q^2 <16$)  & 0.2  & 0.2 \\
Form-factor~model dependence ($q^2 > 16$)       & 0.1  & 2.2 \\
\hline
MC statistics   ($q^2 < 8$)                    & 18.3  & 19.8 \\
MC statistics   ($8 <q^2 <16$)                   & 11.8  & 14.7 \\
MC statistics   ($q^2 > 16$)             & 17.6  & 23.0 \\
\hline
Total error  ($q^2 < 8$)                    & 21.2  & 21.2 \\
Total error    ($8 < q^2 < 16$)                  & 16.0  & 16.7 \\
Total error    ($q^2 > 16$)           & 20.6  & 24.2 \\
\hline
\end{tabular}
\end{center}
\label{tab:systematics2}
\end{table}

\section{Results for the Branching Fractions and \Vub}
\label{sec:vubextract}

The \bpi and \bpiz decay rates are related to \Vub through a hadronic form-factor. 
With the assumption of massless leptons and isospin symmetry, the differential decay rates are given by:

\begin{equation}
\label{eq:partialGamma}
\frac{d\Gamma(\bpi)}{dq^2} = 2\times \frac{d\Gamma(\bpiz)}{dq^2} =
  \frac{G_F^2\Vub^2}{24\pi^3}|f_+(q^2)|^2p_{\pi}^3,
\end{equation}

\noindent where $p_{\pi}$ is the momentum of the pion in the rest frame of the $B$ meson.
The form-factor (FF) $f_+(q^2)$ has been calculated with different assumptions, which 
predict different $q^2$ spectra.

\begin{figure}[top]
 \begin{centering}
 \epsfig{file=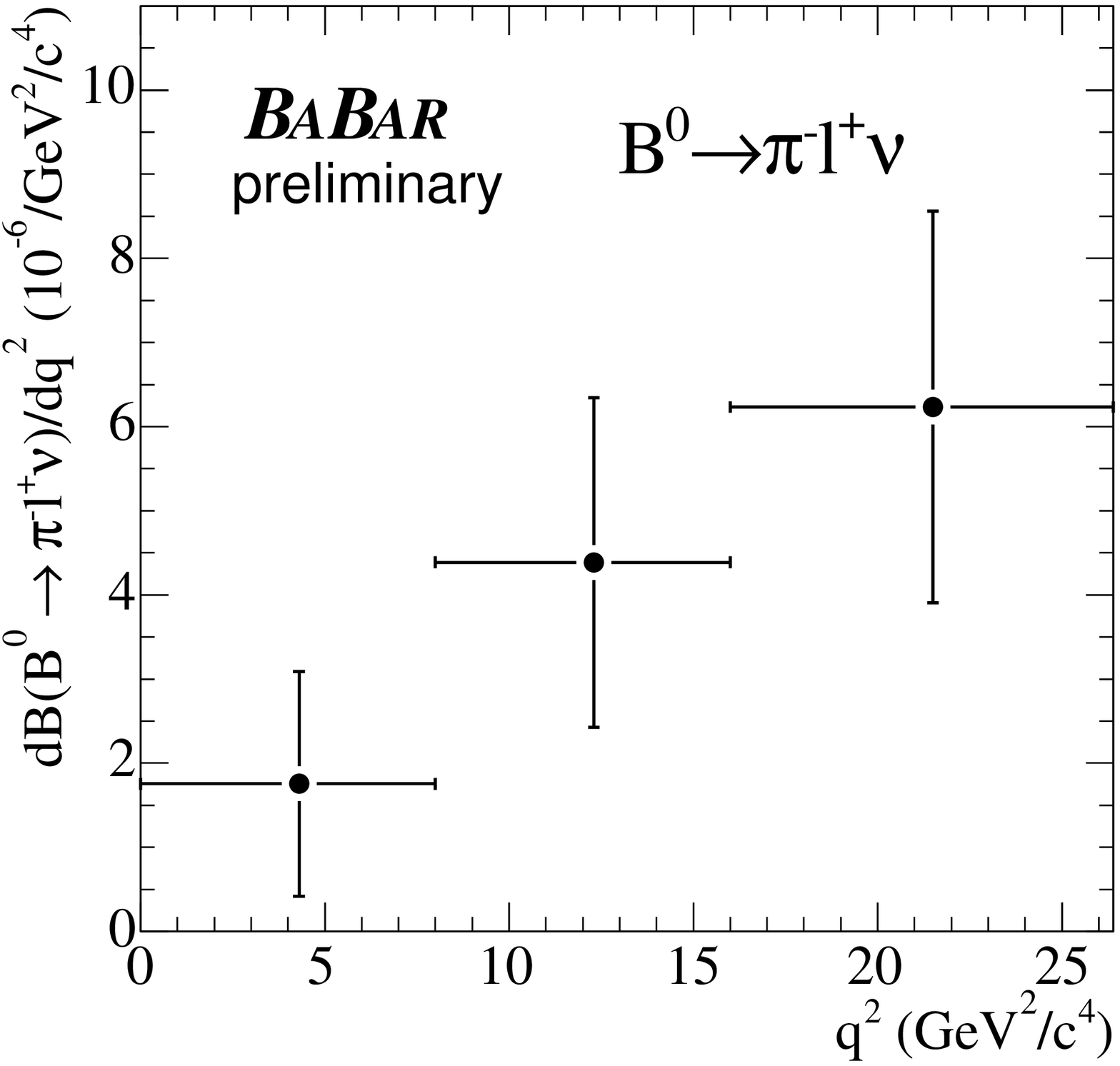,width=8.cm} 
 \epsfig{file=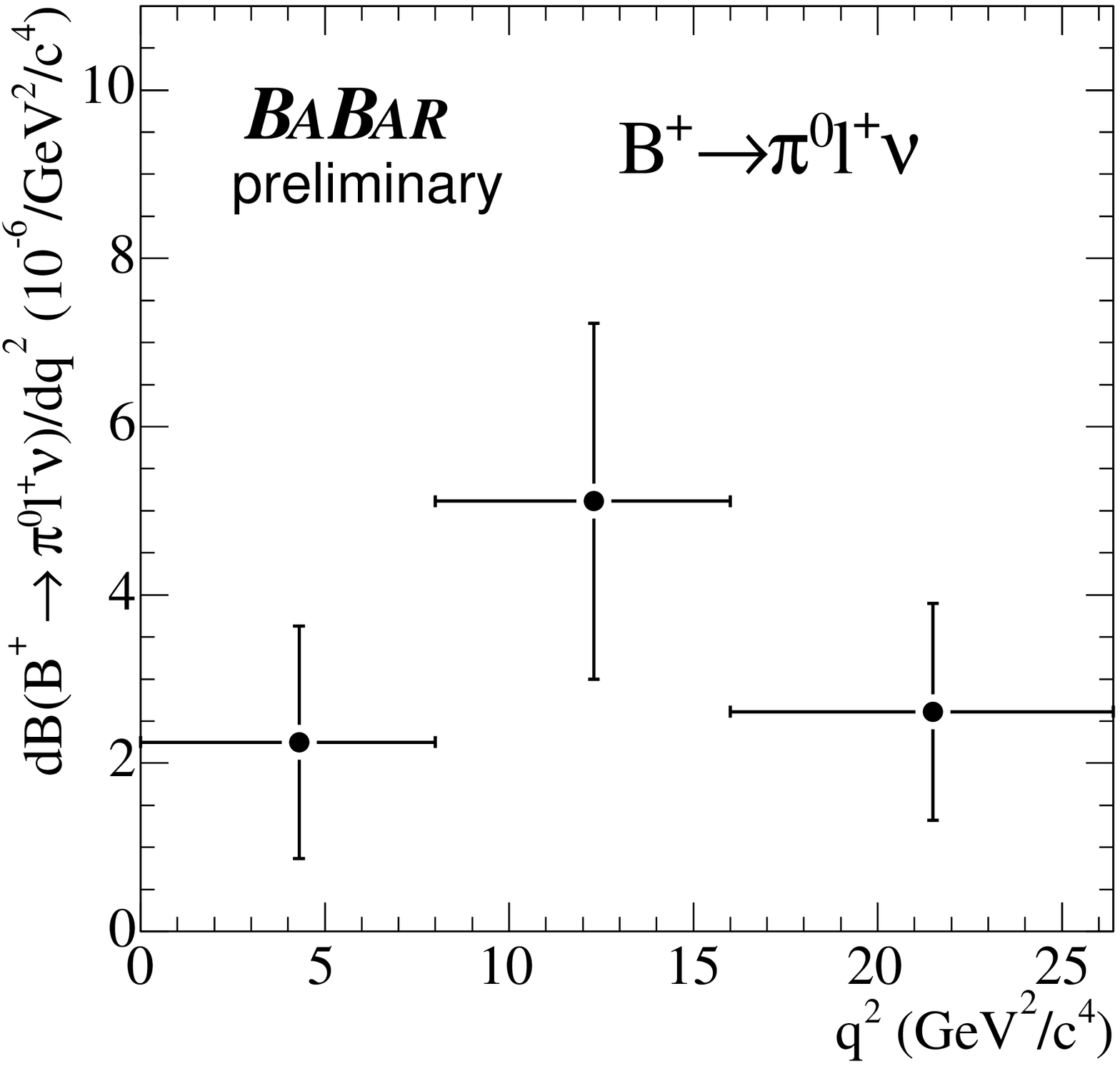,width=8.cm} 
 \epsfig{file=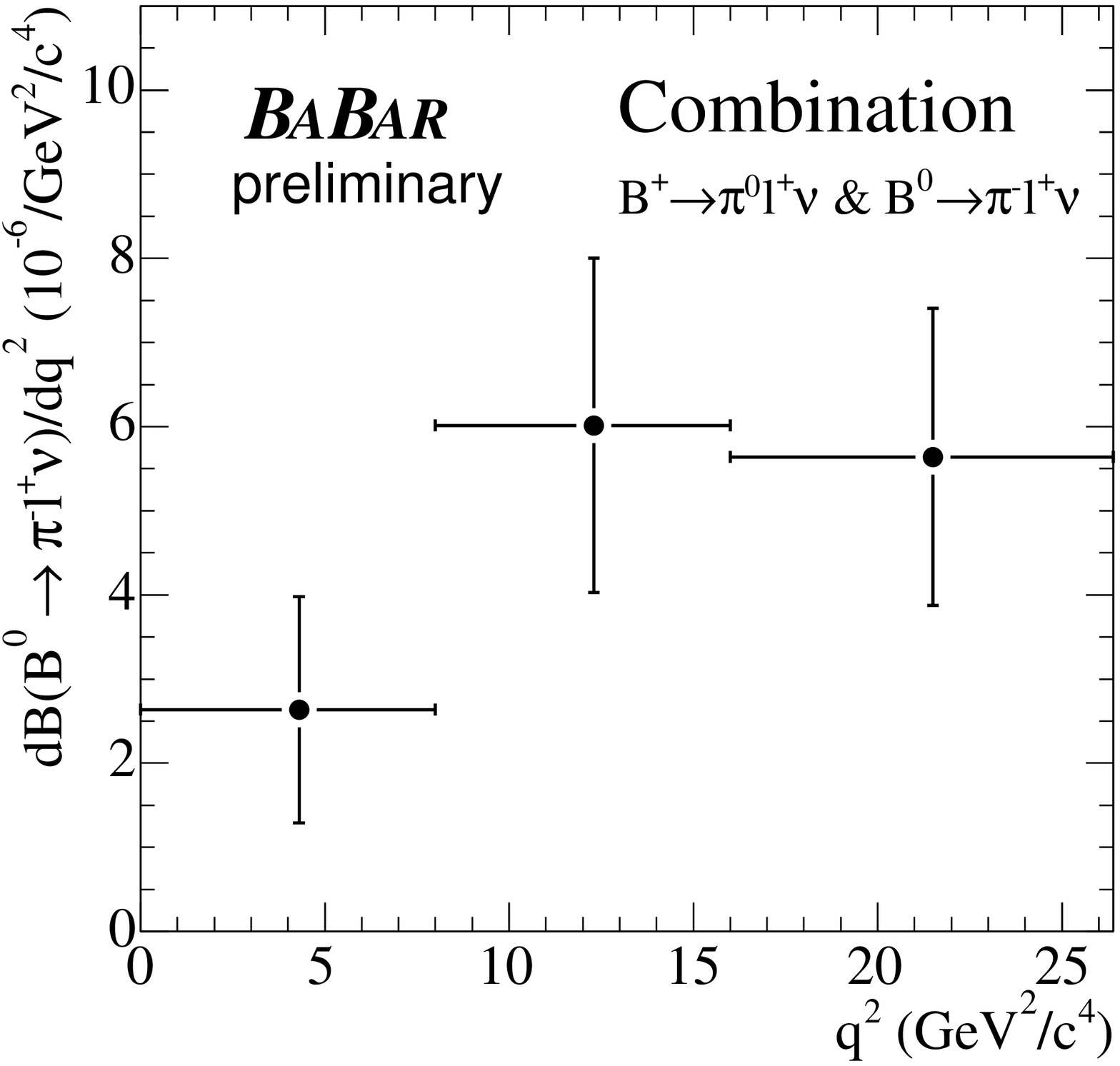,width=8.cm} 
 \epsfig{file=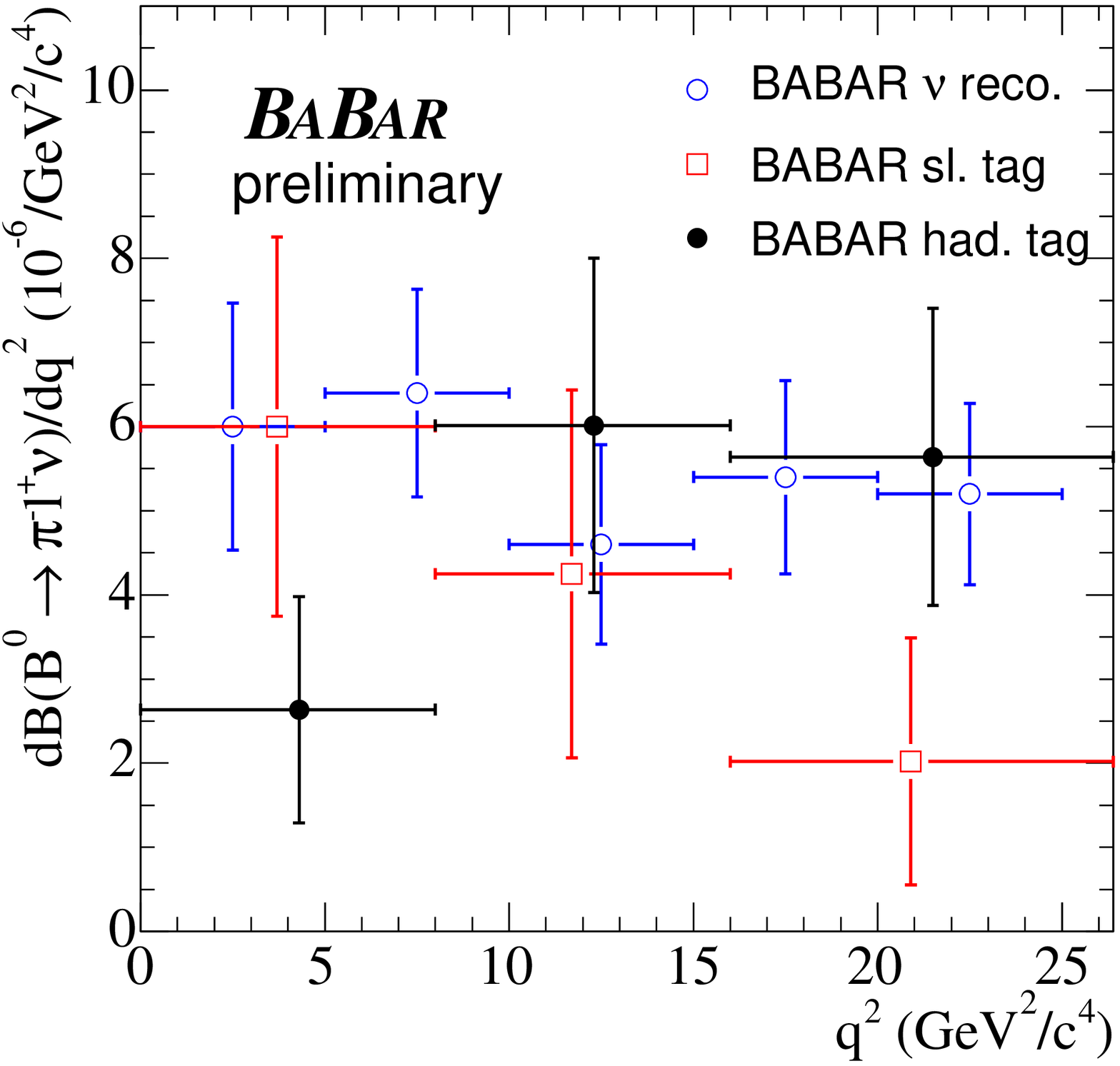,width=8.cm} 
 \caption{ Partial branching ratio in $q^2$ bins for \bpi (top left), \bpiz (top right), combination (bottom left) and comparison of the combination with 
           the previous \babar\ measurements (bottom right): untagged analysis with $\nu$ reconstruction ~\cite{bad1158} and analysis with $D^*\ell\nu$ tag~\cite{piplnu,pizlnu}. The error bars represent the sum in quadrature of statistical and systematic uncertainties.
\label{fig:brq2}}
 \end{centering}
\end{figure}

We derived the partial branching fractions using the inclusive semileptonic branching ratio 
$\BR(\Bxlnu) = (10.73 \pm 0.28)\%$ and the ratio of the \Bz and \Bp lifetimes
$\tau_{B^+}/\tau_{B^0}=1.086 \pm 0.017$ \cite{PDG2004}. The results for the partial branching 
fractions of \bpi and \bpiz are reported in Tab.~\ref{tab:finalresq2pi}.
The partial and total rates for exclusive decays involving \pip and \piz can be constrained
using isospin symmetry, $\Gamma(\bpi) = 2 \times  \Gamma(\bpiz)$, and the ratio of the \Bz and \Bp lifetimes.
We average the six results for the partial branching fractions by following the prescription suggested by the Heavy Flavor Averaging Group in \cite{kowa}. The measurements agree with each other with $\chi^2 = 3.8$ for 3 degrees of freedom.
The results are reported in Tab.~\ref{tab:finalresq2pi} and 
in Fig. \ref{fig:brq2}, including a comparison with other recent \babar\ measurements for partial 
branching ratios.

\begin{table}[top]
\begin{center}
\caption{ Measured partial branching fractions of the  \bpi and  \bpiz decays. The last column lists
the combined measurement for \bpi assuming isospin symmetry. The first error is statistical, 
the second one systematic.}
\vspace{0.1in}
\begin{tabular}{|l|c|c|c|}
\hline
$q^2$ bin                   & $\Delta \BR(\bpi)$ & $\Delta \BR(\bpiz)$  & $\Delta \BR(\bpi)$  \\
 & $[\times10^{-4}]$& $[\times10^{-4}]$& $[\times10^{-4}]$(comb)\\
\hline\hline
$q^2 < 8 \gev^2/c^4$            & $0.14 \pm 0.10 \pm 0.03$& $0.18 \pm 0.10 \pm 0.04$&$ 0.21\pm 0.10 \pm 0.03$\\
$8 < q^2 < 16 \gev^2/c^4$       & $0.35 \pm 0.15 \pm 0.06$& $0.41 \pm 0.16 \pm 0.06$&$ 0.48\pm 0.14 \pm 0.06$\\
$q^2 > 16 \gev^2/c^4$           & $0.65 \pm 0.20 \pm 0.12$& $0.27 \pm 0.12 \pm 0.06$&$ 0.59\pm 0.15 \pm 0.10$\\
\hline			      	 		      
Total                           & $1.14 \pm 0.27 \pm 0.17$& $0.86 \pm 0.22 \pm 0.11$&$ 1.28\pm 0.23 \pm 0.16$\\
\hline 
\end{tabular}
\label{tab:finalresq2pi}
\end{center}
\end{table}

Integrating Eq.~\ref{eq:partialGamma} in a given $q^2$ interval, \Vub\ can be extracted from the measured partial branching fraction $\Delta \BR(\bpi)$ using the following expression:

\begin{equation}
\Vub = \sqrt{\frac{\Delta \BR(\bpi)}{\Delta\zeta\cdot\tau_{\Bz}}},
\end{equation}

\noindent where $\tau_{\Bz} = 1.536\pm0.014\ps$~\cite{PDG2004} is the $\Bz$ lifetime,
$\Delta \BR(\bpi)$ is the combined partial branching fraction for a given $q^2$ interval and $\Delta\zeta$ is the predicted form-factor normalization for the same $q^2$ interval and is defined as:

\begin{equation}
\Delta\zeta = \frac{G_F^2}{24\pi^3}\int_{q^2_{min}}^{q^2_{max}}|f_+(q^2)|^2p_{\pi}^3dq^2.
\end{equation}

\noindent Different theoretical calculations of $f_+(q^2)$ are available in the literature \cite{Ball05,HPQCD04,FNAL04}, which predict different $q^2$ spectra and are considered reliable only in different limited $q^2$ ranges.

To minimize the theoretical error on \Vub the range of $q^2$ therefore is chosen to correspond to the region where the form-factor normalization is considered more reliable: $q^2<16\gev^2/c^4$ for LCSR and $q^2>16\gev^2/c^4$ for unquenched lattice QCD (LQCD) calculations (HPQCD and FNAL in Tab. \ref{tab:vubresult}). The extrapolation of the form-factors to the full $q^2$ range allows for the extraction of \Vub\ from the total branching fraction, but introduces additional uncertainties which must be taken into account.
The calculation of the form-factor over the full $q^2$ range is done in Refs.~\cite{Ball05,HPQCD04,FNAL04} using empirical functions and additional uncertainties are quoted for the extrapolation.
Table~\ref{tab:vubresult} summarizes the values of \Vub\ extracted from the measured partial 
and total branching fractions.

Instead of averaging results based on different theoretical calculations,
we report the value of $\Vub$ obtained from the total branching fraction
based on one of the LQCD calculations~\cite{FNAL04},
\[
\Vub = (3.7 \pm 0.3_{stat} \pm 0.2_{syst} \pm {{}^{+0.8}_{-0.5}}_{FF})\times10^{-3}.
\]
as a representative result, where the last error is due to the normalization of the form-factor. This result is intermediate between the results based on the other two calculations and includes the most conservative estimation of the 
theoretical uncertainty due to the form-factor normalization.

\begin{table}[thbp]
\begin{center}
\caption{\label{tab:vubresult}
Preliminary results of \Vub\ extracted from the measured partial (first three rows) and total 
(last three rows) branching fractions and form-factor calculations.}
\vspace{0.1in}
\begin{tabular}{lccc}
\hline\hline
 FF calculation & $q^2$ range & $\Delta\zeta$ ($\ps^{-1}$) & \Vub ($10^{-3}$) \\
 \hline
 Ball-Zwicky~\cite{Ball05}
 & $<16\gev^2/c^4$ & $5.44\pm1.43$ & $2.9 \pm 0.5_{stat} \pm 0.1_{syst} \pm {{}^{+0.5}_{-0.3}}_{FF}$\\
 HPQCD~\cite{HPQCD04}
 & $>16\gev^2/c^4$ & $1.29\pm0.32$ & $5.4 \pm 0.7_{stat} \pm 0.5_{syst} \pm {{}^{+0.8}_{-0.6}}_{FF}$\\
 FNAL~\cite{FNAL04}
 & $>16\gev^2/c^4$ & $1.83\pm0.50$ & $4.6 \pm 0.6_{stat} \pm 0.4_{syst} \pm {{}^{+0.8}_{-0.5}}_{FF}$\\
 \hline
 Ball-Zwicky~\cite{Ball05}
 & full & $7.74\pm2.32$        & $3.3 \pm 0.3_{stat} \pm 0.2_{syst} \pm {{}^{+0.6}_{-0.4}}_{FF}$\\
 HPQCD~\cite{HPQCD04}
 & full & $5.70\pm1.71$        & $3.8 \pm 0.3_{stat} \pm 0.2_{syst} \pm {{}^{+0.7}_{-0.5}}_{FF}$\\
 FNAL~\cite{FNAL04}
 & full & $6.24\pm2.12$        & $3.7 \pm 0.3_{stat} \pm 0.2_{syst} \pm {{}^{+0.8}_{-0.5}}_{FF}$\\
 \hline\hline
 \end{tabular}
\end{center}
\end{table}

\section{Conclusions}

Using events tagged by a fully reconstructed hadronic decay of one $B$ meson, we have measured
the total branching fractions for \bpi and \bpiz decays:

\[
\BR(\bpi) = (1.14 \pm 0.27_{stat} \pm 0.17_{syst})\times 10^{-4},
\]
\[
\BR(\bpiz) =(0.86 \pm 0.22_{stat} \pm 0.11_{syst})\times 10^{-4}.
\]

\noindent Combining the results assuming isospin symmetry we have extracted the total branching fraction:

\[
\BR(\bpi) = (1.28 \pm 0.23_{stat} \pm 0.16_{syst})\times 10^{-4},
\]

\noindent and the partial branching fractions in three $q^2$ bins:

\[
\Delta\BR(\bpigen)
= \left\{\begin{array}{ll}
  (0.21 \pm 0.10_{stat} \pm 0.03_{syst})\times 10^{-4} & q^2 < 8\gev^2/c^4,\\
  (0.48 \pm 0.14_{stat} \pm 0.06_{syst})\times 10^{-4} & 8 < q^2 < 16\gev^2/c^4,\\
  (0.59 \pm 0.15_{stat} \pm 0.10_{syst})\times 10^{-4} & q^2 > 16\gev^2/c^4.
  \end{array}\right.
\]

\noindent From the measured partial branching fractions and the theoretical predictions for the form-factor
normalization we obtained a preliminary determination of the CKM matrix element \Vub:

\[
\Vub = (3.7 \pm 0.3_{stat} \pm 0.2_{syst} {{}^{+0.8}_{-0.5}}_{FF})\times10^{-3},
\]

\noindent where the last error is due to the form-factor normalization.

\section{ACKNOWLEDGMENTS}
\label{sec:Acknowledgments}
We are grateful for the 
extraordinary contributions of our \pep2\ colleagues in
achieving the excellent luminosity and machine conditions
that have made this work possible.
The success of this project also relies critically on the 
expertise and dedication of the computing organizations that 
support \babar.
The collaborating institutions wish to thank 
SLAC for its support and the kind hospitality extended to them. 
This work is supported by the
US Department of Energy
and National Science Foundation, the
Natural Sciences and Engineering Research Council (Canada),
Institute of High Energy Physics (China), the
Commissariat \`a l'Energie Atomique and
Institut National de Physique Nucl\'eaire et de Physique des Particules
(France), the
Bundesministerium f\"ur Bildung und Forschung and
Deutsche Forschungsgemeinschaft
(Germany), the
Istituto Nazionale di Fisica Nucleare (Italy),
the Foundation for Fundamental Research on Matter (The Netherlands),
the Research Council of Norway, the
Ministry of Science and Technology of the Russian Federation, and the
Particle Physics and Astronomy Research Council (United Kingdom). 
Individuals have received support from 
CONACyT (Mexico),
the A. P. Sloan Foundation, 
the Research Corporation,
and the Alexander von Humboldt Foundation.

\cleardoublepage

\end{document}